\documentclass[bibyear]{aa} 

\usepackage{graphicx}
\usepackage{txfonts}
\usepackage{threeparttablex}
\usepackage{lscape} 
\usepackage{xcolor, soul}
\sethlcolor{yellow}
\usepackage{geometry}
\usepackage{longtable}
\usepackage{natbib}
\usepackage{hyperref} 
\bibpunct{(}{)}{;}{a}{}{,}             

\begin{document} 

   \title{Tidally driven tectonic activity as a \\ parameter in exoplanet habitability}

  \author{S.R.N. McIntyre
          \inst{1,2}
          }
   \institute{Research School of Astronomy and Astrophysics, Australian National University, Canberra, ACT 2601, Australia\
   \
   \and
        Research School of Earth Sciences, Australian National University, Canberra, ACT 2601, Australia\\
                \email{sarah.mcintyre@anu.edu.au}
        }


 
  \abstract
   {Habitability of an exoplanet is defined by its ability to support surface liquid water. The long-term carbon cycle plays an important role in regulating planetary temperature, thus aiding the conditions for the preservation of surface liquid water and, consequently, the habitability of exoplanets.}
   {On Earth, plate tectonics play an integral role in driving the long-term carbon cycle; however, on tidally locked rocky exoplanets alternative tectonic mechanisms driven by tidal stress and tidal heating could serve in an analogous way.}
   {We calculate tidal stress and tidal heating rates to model the likelihood of tectonic activity maintaining stable climates suitable for surface liquid water on tidally locked rocky exoplanets with radii ${R}_{p}$ $\le$ 1.23R$_\oplus$.}
   {Applying the tidal models to our sample of 767 tidally locked rocky exoplanets reveals that $\sim$10\% of exoplanets, including Proxima Cen b and GJ 1061 d from the circumstellar habitable zone (CHZ), pass the tidal stress subduction threshold for mobile lid tectonic activity \textit{and} reside within the optimal tidal heating zone. This subset of exoplanets could sustain tidally induced temperate mobile lid tectonic activity comparable to plate tectonics on Earth, aiding in maintaining the presence of surface liquid water. Furthermore,  $\sim$40\% of exoplanets from our sample located in the CHZ would be unable to maintain the tectonic activity needed to stabilise the climate and are unlikely to retain surface liquid water. When broadening our modelling to establish the overlap between tidal stress, tidal heating, and the CHZ, to discover optimal regions to target for future observations, we determine that tidally driven tectonic activity conducive to the maintenance of surface liquid water occurs predominantly around M dwarfs, and identify intersections, where both mobile lid and optimal tidal heating could be sustained on eccentric (\textit{e}>0.1) Earth-sized exoplanets (${R}_{p}$ = 1.0-1.23R$_\oplus$) orbiting in the CHZ of low-mass M dwarfs.}
   {}


   \maketitle
%

\section{Introduction} 
\label{sec:intro}

Our universe contains approximately 10${}^{22}$ stars, with $\sim$10${}^{21}$ of them like our Sun, many of which host rocky exoplanets orbiting in their habitable zone  \citep{mottl2007water,pizzarello2007chemistry, robles2008comprehensive, kopparapu2013habitable,kaltenegger2017characterize}. More than 5000\footnote{\url{https://exoplanetarchive.ipac.caltech.edu/cgi-bin/TblView/nph-tblView?app=ExoTbls&config=PSCompPars}-(accessed 28 February 2022) \label{nasa0}} planets have been detected over the past 28 years, and their diversity has broadened our understanding of exoplanetary environments. The classical definition of habitability, involving rocky bodies with enough surface gravity to sustain an atmosphere orbiting their host stars at a distance where stellar radiation is suitable for the presence of surface liquid water, fails to account for a multitude of factors that impact a planet’s ability to support liquid water and conditions favourable for the evolution of life \citep{kaltenegger2017characterize}. Recent research shows that the conservation of liquid water on exoplanets depends on a diverse range of astronomical and planetary parameters such as tectonic activity, magnetic field, stellar type, albedo, planet chemical composition, tidal locking, and impact events \citep{cockell2016habitability,noack2017volcanism,rushby2018long,gallet2019impact,mcintyre2019planetary,yamashiki2019impact,chambers2020effect,herath2020characterizing}. In order to determine optimal targets for future biosignature observations, it is increasingly important to expand our research into further defining these habitability factors. 

Preservation of surface liquid water requires the stabilisation of climate through the long-term carbon cycle, so the planet is neither uninhabitably cold or hot due to variations in greenhouse gas concentrations or stellar luminosity \citep{walker1981negative,berner1997need,foley2016whole,noack2017volcanism,valencia2018habitability,foley2018carbon,foley2019habitability,van2019exoplanet}. On Earth, plate tectonics play an integral role in regulating temperature and are often posited to be essential in supporting conditions for a long-lived habitable climate and thus long-term surface liquid water \citep{kasting2003evolution,tikoo2017fate}. Through the subduction of rocks over large areas, plate tectonics provide a pathway for CO${}_{2}$ isolated in carbonate rocks to return to the atmosphere \citep{cockell2016habitability}. The negative feedback inherent in this path enables the carbon cycle to act as a long-term thermostat and regulate Earth's surface temperature to remain within the scope suitable for liquid water \citep{walker1981n,rushby2018long}.

Venus and Mars demonstrate opposing consequences of the lack of this cycle. On Venus, the loss of liquid water due to the greenhouse effect inhibits carbonate formation, with any buried carbon having long since been heated and returned to the planet's thick CO${}_{2}$ atmosphere \citep{taylor2009climate}. Alternatively, in the case of Mars, a lack of continuous volcanic outgassing is suspected of having led to an atmosphere too thin to allow a substantial greenhouse effect and consequently conditions required to maintain surface liquid water \citep{lammer2013outgassing}. Earth appears to be within an optimal regime for the maintenance of the long-term carbon cycle, which has assisted in the preservation of a stable surface temperature required for liquid water \citep{lunine2013earth}.
The process of plate tectonics on Earth is sustained by internal heating driving convection in the mantle, suggesting that a similar source of heating is required to initiate and maintain tectonic activity on extrasolar planets \citep{o2007conditions}. For Earth, a combination of the primordial heat remaining from the planet's formation and the decay of radionuclides U and K provides an adequate internal heating budget to initiate plate tectonics \citep{o2016window}. However, rates for primordial and radiogenic heating of exoplanets are unknown and might be inadequate to drive long-lived tectonic activity \citep{barnes2009tidal}. \citet{jackson2008tidal} and \citet{zanazzi2019ability} show that tides could be a sufficient source of internal heating and tidal stress required to initiate and maintain tectonic activity on tidally locked rocky exoplanets, which comprise $\sim$99\% of currently observed exoplanets \citep{mcintyre2019planetary}.
Until recently, it was thought that plate tectonics were necessary for the long-term carbon cycle to operate and stabilize climates. However, there is now a wealth of literature demonstrating that other tectonic mechanisms are able to initiate and maintain this cycle. \citet{tosi2017habitability} and \citet{dorn2018outgassing} calculate outgassing rates for stagnant lid planets and show long-lived CO${}_{2}$ outgassing can occur; \citet{foley2018carbon} and \citet{foley2019habitability} demonstrate that silicate weathering can still operate without plate tectonics and calculate conditions where habitable climates can persist on such planets; while \citet{valencia2018habitability} outline how the long-term carbon cycle could operate on temperate stagnant lid exoplanets by vertically recycling CO${}_{2}$ between the atmosphere, basaltic crust, and mantle through tidally induced volcanism, sea weathering, basalt formation, and foundering.
Here, we focus on modelling the effects of tidal stress and tidal heating on tidally locked rocky exoplanets, with radii ${R}_{p}$ $\le$ 1.23 R$_\oplus$, to assess their impact on the exoplanets’ ability to sustain a long-term carbon cycle, enabling conditions suitable for the presence of surface liquid water.
\section{Model} 
\label{sec:modl}
\subsection{Tidal stress model} 
\label{sec:tidstress}

One of the main distinctions between the different tectonic scenarios involves determining whether the planet’s lithosphere is in a mobile lid convection regime or a stagnant lid regime (singular rigid plate covering the convecting mantle) \citep{bercovici20157,lenardic2018diversity,stevenson2019planetary,zanazzi2019ability}. 

A lateral force pushing one lithospheric plate under another at a fault line, the process of subduction, is required for a planet to develop a mobile lid tectonic regime. This lateral force needs to overcome the frictional force between the two plates. On Earth, stresses operating on the lithosphere from mantle convection drive the main lateral force \citep{bercovici20157}. Under certain conditions, tidal stresses are also able to apply a lateral force on lithospheric plates and could initiate subduction on exoplanets if sufficiently strong \citep{zanazzi2019ability}. 

To determine the strength of tidal stress for tidally locked rocky exoplanets, we use \citet{zanazzi2019ability} equation for eccentric ultra-short period planets:
\begin{equation} 
\label{eq:subeq1} 
{h} = \frac{3.8 \times {{10}^{-4}} }{1 + {\overline{\mu}}} \left(\frac{M_ *}{1M_\odot}\right) {\left(\frac{M_p}{5M_\oplus}\right)}^{-1} {\left(\frac{R_p}{1.6R_\oplus}\right)}^3 \left(\frac{e}{0.01}\right) {\left(\frac{a}{0.015AU}\right)}^{-3} \end{equation}
\noindent where ${R}_{p}$ and ${M}_{p}$ are the planet’s radius and mass, respectively, ${M}_{*}$ is the stellar mass, \textit{e} is the planet's eccentricity, and ${a}$ is the planet’s semi-major axis. Furthermore, we follow \citet{zanazzi2019ability} assumption of $\mu = 10^{12}$ dynes/$cm^2$ and compute the re-scaled rigidity $\overline{\mu}$ as:
\begin{equation} 
\label{eq:ridgeq} 
{\overline{\mu}} = \frac{38{\pi}{\mu}{{R_p}^4}}{{3}{G}{{M_p}^2}}  \end{equation}

According to \citet{zanazzi2019ability} the lateral force from tidal stresses acting on a lithospheric plate can be used to calculate a critical tidal bulge, where ${h} > {10^{-5}}$ is enough convective stress to aid or initiate subduction. Thus, we use this tidal stress threshold when applying Equation \ref{eq:subeq1} to a sample of rocky exoplanets. This model of tidal stress assumes a homogeneous interior composition, constant density incompressible elastic solid planet, with a uniform shear modulus and viscosity modelled based on rocky planets observed in the solar system.

\subsection{Tidal heating model} 
\label{sec:tidheat}

Tectonic activity driven by tidal heating on tidally locked planets, has been suggested to be analogous to plate tectonics on Earth in assisting a long-term carbon cycle \citep{jackson2008tidal,valencia2018habitability}. The orbital energy of planets on eccentric orbits dissipates as tidal energy in the interior, drives inward migration and circularises the planet’s orbit, resulting in significant internal heating \citep{driscoll2015tidal}. 

We use the tidal heating model as collated by \citet{jackson2008tidal}:
\begin{equation} 
\label{eq:tidale} 
\frac{1}{e}\frac{de}{dt}=- \left[\frac{63}{4}{\left(G{M_*}^3\right)}^{1/2}\frac{{R_p}^5}{Q'_pM_p}+\frac{171}{16}{\left(G/M_*\right)}^{1/2}\frac{{R_*}^5M_p}{Q'_* }\right]{a}^{-13/2}
\end{equation}
\begin{equation} 
\label{eq:tidala} 
\frac{1}{a}\frac{da}{dt}=- \left[\frac{63}{2}{\left(G{M_*}^3\right)}^{1/2}\frac{{R_p}^5}{Q'_pM_p}{e}^2+\frac{9}{2}{\left(G/M_*\right)}^{1/2}\frac{{R_*}^5M_p}{Q'_* }\right]{a}^{-13/2}
\end{equation}

\noindent where $Q'_*$  and $Q'_p$ are stellar and planetary dissipation, respectively. For the stellar dissipation factor, we follow \citet{jackson2008tidal} assumption of $Q'_* = 10^{5.5}$. We iterate $Q'_p$ over a range of values for the planetary dissipation factor, $Q'_p$ = 10 to 500, comparable with the variation found for the terrestrial planets and satellites in the solar system \citep{goldreich1966q}. 

\citet{barnes2009tidal} conclude that the relationship between tidal heating rate and loss of orbital energy is proportional to the initial term in Equation~\ref{eq:tidale}. Thus, we can express the tidal heating rate (\textit{H}) as:
\begin{equation} 
\label{eq:tidalh} 
{H}= \frac{63}{4}\frac{{\left(G{M_*}\right)}^{3/2}M_*{R_p}^5}{Q'_p}{a}^{-15/2}{e}^2
\end{equation}

\subsection{Sample selection and Monte Carlo calculations} 
\label{sec:sampsel}

We utilise NASA's composite planet database\footnote{\url{https://exoplanetarchive.ipac.caltech.edu/cgi-bin/TblView/nph-tblView?app=ExoTbls&config=compositepars} (accessed 28 February 2022) \label{nasa}}, in combination with additional information on the Kepler planets' radii provided by \citet{berger2020gaiacat1} and updated stellar properties by \citet{berger2020gaiacat1} to compose a catalogue of tidally locked rocky exoplanets (Table~\ref{tab:exostres}). 

Furthermore, we use the \citet{chen2016probabilistic} M-R relationship to calculate unknown radii or masses and their uncertainties:
\begin{equation} 
\label{eq:masseq} 
{R_p} \sim {{M_p}^\mathrm{0.279 \pm 0.009}} 
\end{equation}

To ensure we include only rocky exoplanets in our database, we follow the \citet{chen2016probabilistic} definition of the boundary between terrestrial and Jovian planets, limiting our selection to exoplanets with radii ${R}_{p}$ $\le$ 1.23 R$_\oplus$. 

\citet{jackson2008tidal} tidal heating model does not include heating from non-synchronous rotation and non-zero obliquity. Thus, for the tidal heating model to work, each planet needs to be tidally locked, with a
rotation that had spun down early enough in the planet's history that the interchange of angular momentum between the planet's rotation and its orbit is not a significant factor  \citep{jackson2008tidal}. To ensure we are only examining tidally locked planets, we use \citet{griessmeier2009protection} equation to estimate the time required to tidally lock each planet in our sample: 

\begin{equation} \label{eq:rotp}
{\tau} _{sync}\approx \frac{4}{9}\alpha Q'_p\left(\frac{{R_p}^3}{GM_p}\right)\left({\mathit{\Omega}}_i-{\mathit{\Omega}}_f\right){\left(\frac{M_p}{M_ *}\right)}^2{\left(\frac{a}{R_p}\right)}^6 \end{equation}

\noindent where we set $\alpha$ = 1/3, according to Earth's structure parameter. We iterate the ${\tau }_{sync}$ model over a range of values for the planetary dissipation factor, ${Q'_p}$ = 10 to 500, comparable with the variation found for the terrestrial planets and satellites in the solar system \citep{goldreich1966q}. For a realistic approximation, we use an initial rotation rate of $\Omega_i = 5.86\Omega_\oplus$, analogous to early Earth's 4 hr rotation period before the moon-forming impact \citep{canup2008accretion}, and use the current rotational angular velocity $\Omega$ for the final rotation rate, which can be extrapolated from the planet’s orbital period ($\Omega = \frac{2\pi}{P}$). These ${\tau }_{sync}$ values are compared with the stellar ages of the host stars (Table~\ref{tab:exostres}) to ensure only rocky exoplanets that are tidally locked (${\tau} _{sync}$ $\le$ stellar age) are included in our sample. For stars that do not have an age estimate, we assume a lower limit of 1 Gyr. 

Our total sample contains 767 tidally locked rocky exoplanets, with a subset of 14 planets inhabiting the circumstellar habitable zone (CHZ) of their host stars - where the CHZ is the insolation between recent Venus and ancient Mars as defined by the \citet{kopparapu2013habitable} optimistic habitable zone \citep{kopparapu2013habitable,kopparapu2014habitable}. 

For each exoplanet in our sample, we execute 10,000 Monte Carlo simulations that include uncertainties on the stellar masses, stellar radii, planetary masses, planetary radii, planetary dissipation, semi-major axis, eccentricity, and orbital period as provided by the NASA composite planet database and \citet{berger2020gaiacat1}. To ensure that the uncertainties are correctly correlated, we include Equation \ref{eq:masseq} in the Monte Carlo simulations for the subgroup of exoplanets that employ the M-R relation. These simulations allow us to determine the median and 68\% confidence intervals on tidal stress ${h}$ and tidal heating ${H}$ rates.

\section{Data analysis} \label{sec:datany}

\begin{figure}[ht]
    \centering
	\includegraphics[width={250pt}, height={110pt}]{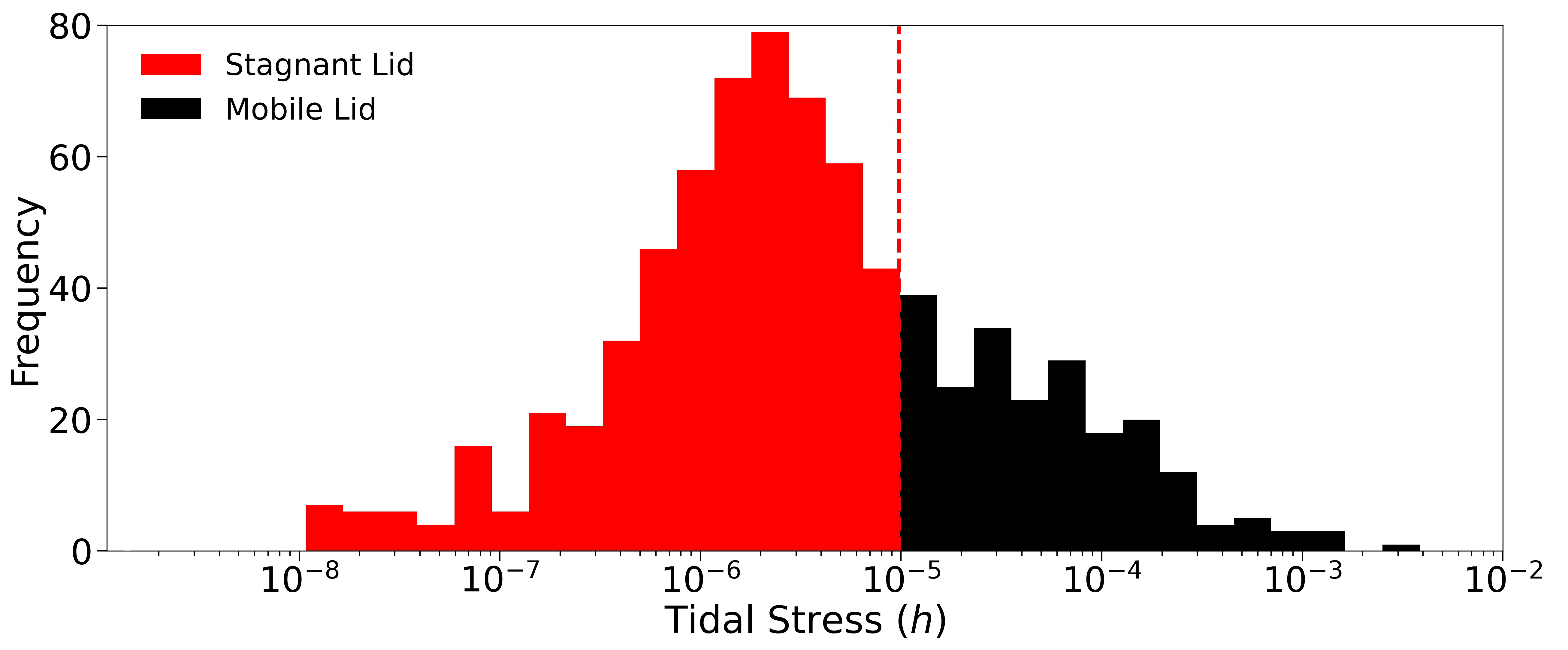}
    \caption{Frequency distribution of the tidal stress values for the 767 exoplanets from our sample. The dashed red line indicates the threshold when a rocky exoplanet experiences sufficient tidal stress (${h} > {10^{-5}}$) to aid or initiate subduction. For exoplanets without an eccentricity value or where \textit{e}$\sim$0, we use \citet{zanazzi2019ability} assumption of \textit{e} = 0.01 $\pm$ 0.01 as reasonable for ultra-short period exoplanets.}  
    \label{fig:figure1}
\end{figure}

When a rocky exoplanet experiences significant tidal stress (${h} > {10^{-5}}$), these tidal stresses can aid mantle convective stress in subducting plates or can even completely initiate subduction over the surface of the exoplanet \citep{zanazzi2019ability}. The $h$ values (see Table~\ref{tab:exostres}) for all exoplanets in our sample were computed using Equations \ref{eq:subeq1} - \ref{eq:ridgeq}  and plotted in Figure~\ref{fig:figure1}. Out of the sample of 767 tidally locked rocky exoplanets, 28\% $\pm$ 1\% pass the subduction threshold. Therefore, these exoplanets have significant tidal stresses to apply a lateral force on lithospheric plates and are more likely to have a mobile lid tectonic regime. Conversely, the 72\% $\pm$ 1\% of exoplanets from our sample that do not have sufficient tidal stresses to pass the subduction threshold, fall within the stagnant lid regime. Figure~\ref{fig:figure1} indicates that the majority of exoplanets do not cross the subduction threshold, implying that our current observational exoplanet data has a bias towards stagnant lid planets. 

\begin{figure}[ht]
    \centering
	\includegraphics[width={250pt}, height={230pt}]{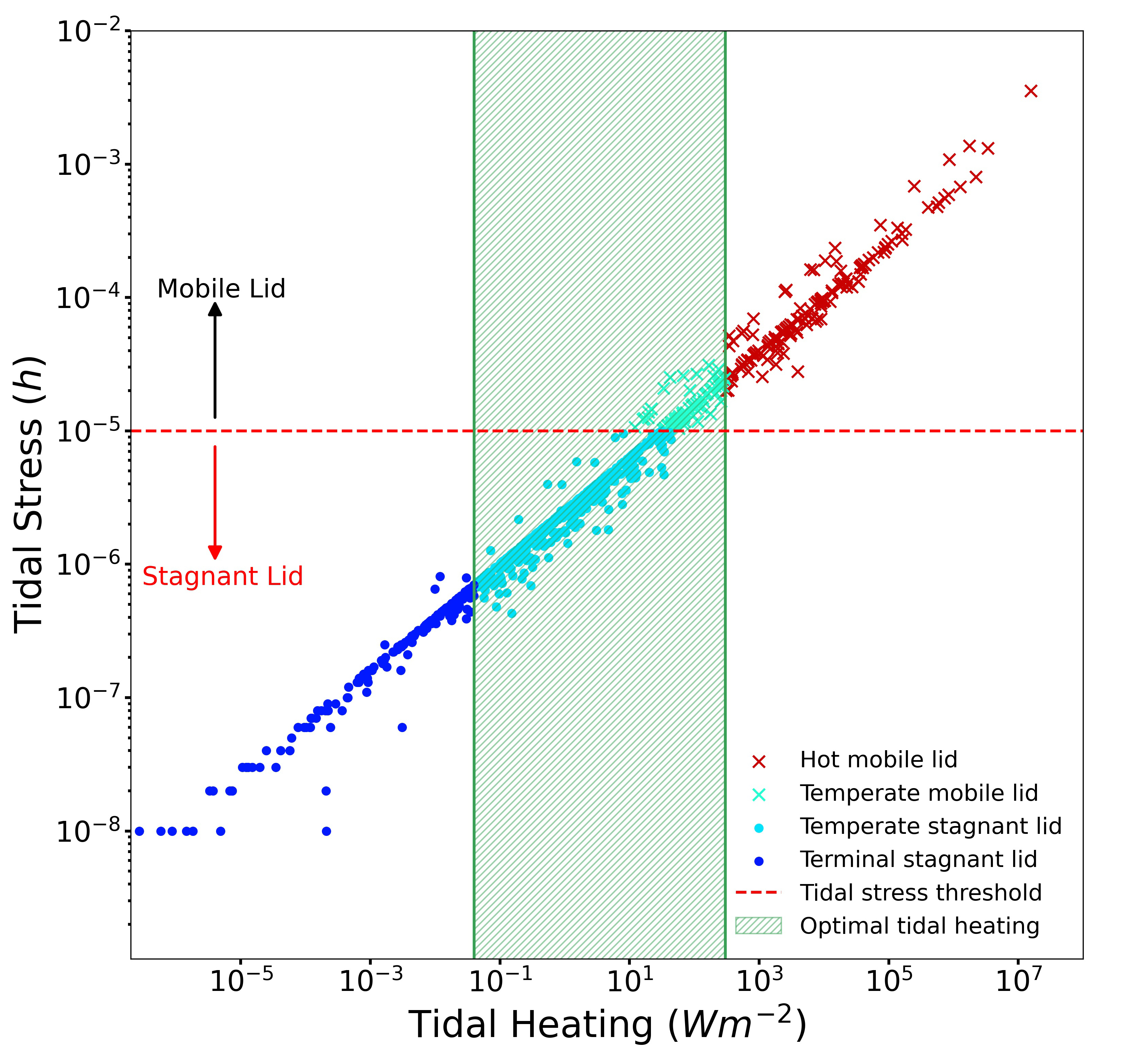}
    \caption{Tidal heating rates and tidal stress values for the 767 exoplanets in our sample. The horizontal dashed red line indicates the threshold when rocky exoplanets experience sufficient tidal stress (${h} > {10^{-5}}$) to aid or initiate subduction. The vertical green shaded region denotes the optimal tidal heating regime between 0.04 and 300 $Wm^{-2}$.} 
    \label{fig:figure2}
\end{figure}

After determining which exoplanets have mobile vs stagnant lid regime, we compute the tidal heating rates using Equation \ref{eq:tidalh} (see Table~\ref{tab:exostres}) and subsequently plot the results in Figure~\ref{fig:figure2}. There is no well-established relationship between internal heating rates and habitability, so to facilitate the interpretation of our modelled ${H}$-values, we contextualise our values with two solar system scenarios at opposite ends of the tectonic regime spectrum. 

For an assessment of the minimum internal heating required, we consider the geologic history of Mars before the termination of tectonic activity resulted in limited volatile cycling between the interior and surface, and CO${}_{2}$ no longer recycling through subduction \citep{ehlmann2016sustainability}. Martian geophysical research highlights a correlation between the termination of tectonic activity and the reduction in internal heating flux below $\sim$0.04 $Wm^{-2}$ \citep{williams1997habitable}. We will, therefore, assume that the exoplanets with tidal heating rates below 0.04 $Wm^{-2}$ fall within the terminal stagnant lid category and do not have the minimum amount of heating needed to maintain tectonic activity, support a stable climate, and would be unable to sustain surface liquid water via tidally induced tectonic activity alone, unless supplemented with an additional source of heating such as radiogenic or primordial heating \citep{stern2018stagnant,jackson2008tidal}. 

For the upper limit, we examine the inner boundary of the CHZ defined by \citet{kasting1988runaway} as the stellar distance where insolation increases to the point where all water on the planet’s surface and in the atmosphere is removed. \citet{barnes2013tidal} refer to exoplanets that experience tidal heating rates high enough to induce this runaway greenhouse, enabling total hydrogen escape, as “Tidal Venuses”. Due to the relationship between IR opacity, vapor pressure, and temperature, for orbital distances less than 1AU, the outgoing IR flux becomes independent of surface temperature and tends asymptotically towards a runaway greenhouse threshold upper limit of $\sim$300 $Wm^{-2}$ \citep{abe1988evolution,kasting1988runaway,ishiwatari2002numerical,selsis2007habitable,barnes2013tidal}. Consequently, we assume that in the circumstances where internal heating rates rise above 300 $Wm^{-2}$, the atmosphere would be unable to maintain radiation balance, resulting in runaway heating of the surface and evaporation of all surface liquid water.

The application of tidal stress and tidal heating models to our sample of 767 tidally locked rocky exoplanets can be seen in Figure~\ref{fig:figure2}. When comparing tidal heating values against our upper and lower boundaries, 18\% $\pm$ 1\% of the modelled exoplanets have a tidal heating rate above 300 $Wm^{-2}$, indicating a reduced likelihood of the existence of liquid water due to runaway greenhouse effect. All these planets pass the initial subduction threshold, suggesting that they are hot mobile lid planets and subsequently are unlikely to retain surface liquid water, unless supplemented by an impact of a water-rich body that alters the atmosphere in a way to stop the desiccating greenhouse effect \citep{barnes2013tidal}. Additionally, 19\% $\pm$ 2\% of the exoplanets have a tidal heating rate below 0.04 $Wm^{-2}$, and as none of these planets pass the initial subduction threshold, this suggests that they are terminal stagnant lid planets and subsequently are unlikely to retain surface liquid water via tidal heating alone. Furthermore, 63\% $\pm$ 2\% of exoplanets reside within the optimum regime for internal heating. Of these, 53\% $\pm$ 2\% do not pass the subduction threshold and are thus temperate stagnant lid exoplanets. On these exoplanets, additional mechanisms like vertical recycling, silicate weathering, or significant long-lived CO${}_{2}$ outgassing would be required to enable stable climates suitable for the presence of surface liquid water \citep{tosi2017habitability,dorn2018outgassing,valencia2018habitability, foley2018carbon, foley2019habitability}. Finally, 10\% $\pm$ 1\% of exoplanets in our sample reside within the optimal tidal heating regime \textit{and} pass the subduction threshold. This subset of exoplanets could be able to sustain tidally induced temperate mobile lid tectonic activity that would help in maintaining the presence of surface liquid water and could be comparable to plate tectonics on Earth.

\begin{figure*}
    \centering
	\includegraphics[width={500pt}, height={310pt}]{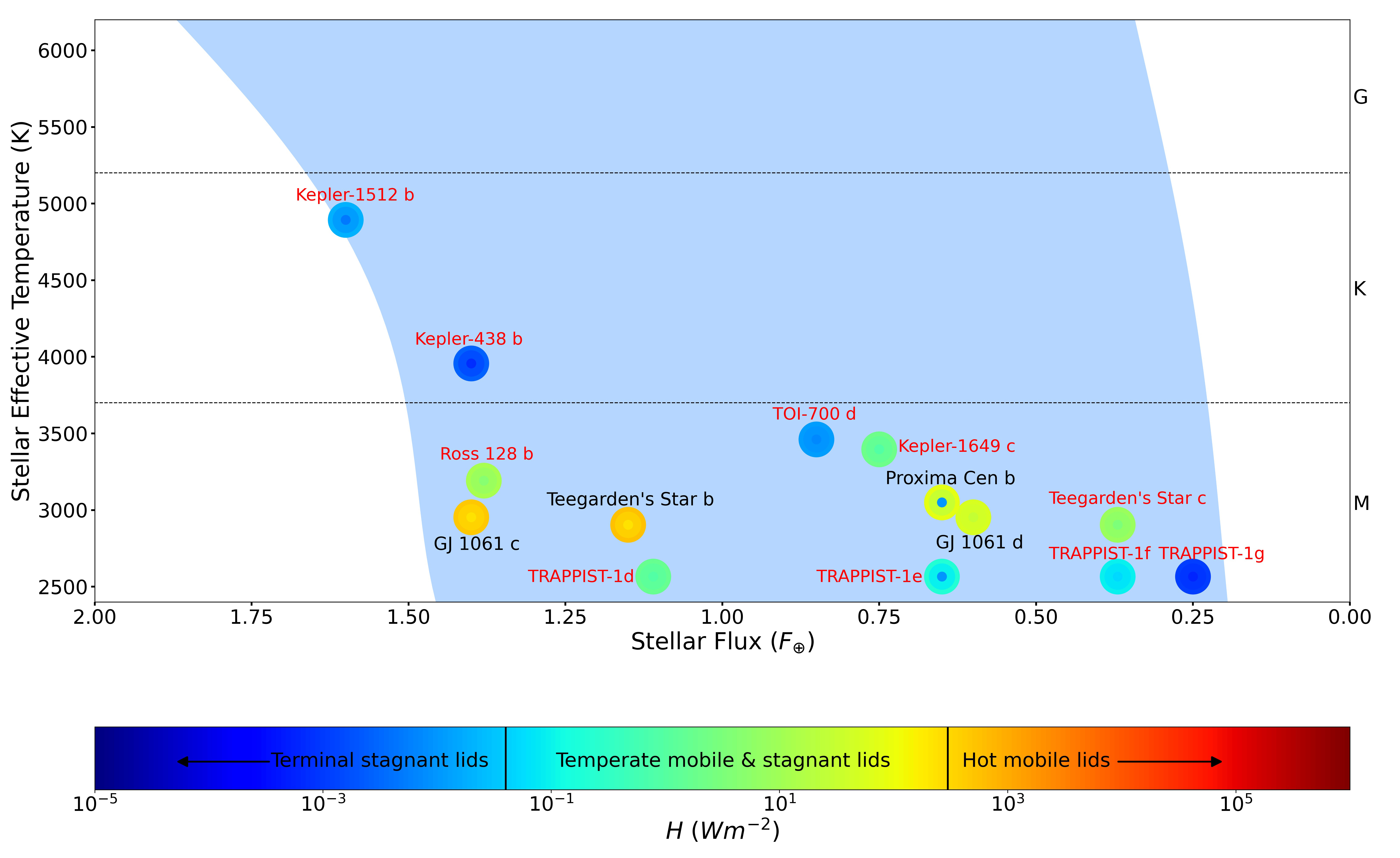}
    \caption{Tidal heating rates of rocky planets located in the CHZ \citep{kopparapu2013habitable,kopparapu2014habitable}. Point colour indicates the planet’s tidal heating values, with upper and lower 68\% confidence values displayed as outer- and inner-most circles, respectively. Labels of stagnant lid exoplanets are red; labels of mobile lid exoplanets are black.} 
    \label{fig:figure3}
\end{figure*}

Focusing on the CHZ exoplanets in Figure~\ref{fig:figure3}, tidal heating rates of Proxima Cen b, GJ 1061 d, TRAPPIST-1d, TRAPPIST-1e, TRAPPIST-1f, Kepler 1649 c, Teegarden's Star c, and Ross 128 b fall within the optimal range between 0.04 and 300 $Wm^{-2}$. Of these temperate exoplanets, Proxima Cen b and GJ 1061 d have subduction values indicative of mobile lid tectonic activity that could help in maintaining the presence of surface liquid water. The remaining temperate CHZ exoplanets TRAPPIST-1d, TRAPPIST-1e, TRAPPIST-1f, Kepler 1649 c, Teegarden's Star c, and Ross 128 b have subduction values indicative of stagnant lid tectonic activity, requiring additional mechanisms like vertical recycling, silicate weathering, or significant long-lived CO${}_{2}$ outgassing, to sustain stable climates suitable for habitability and the presence of  surface  liquid  water \citep{tosi2017habitability,dorn2018outgassing,valencia2018habitability, foley2018carbon, foley2019habitability}. GJ 1061 c and Teegarden's Star b pass the subduction threshold and are likely to have mobile lid tectonics; however, their tidal heating rates are above 300 $Wm^{-2}$, indicating a high probability of greenhouse runaway effect desiccating these hot mobile lid exoplanets. At the opposite end of the spectrum, Kepler 1512b, Kepler 438 b, TOI-700 d, and TRAPPIST-1g do not pass the subduction threshold and have insufficient tidal heating rates to initiate and maintain tectonic activity. These tidally driven terminal stagnant lid exoplanets would be unable to sustain surface liquid water via tidally induced tectonic activity alone; however, with an extra heating source (radiogenic or primordial), \textit{and} additional mechanisms such as vertical recycling, silicate weathering, or significant long-lived CO${}_{2}$ outgassing, these exoplanets could maintain stable climates suitable for the preservation of surface liquid water \citep{tosi2017habitability,dorn2018outgassing,valencia2018habitability, foley2018carbon, foley2019habitability}.

\begin{figure*}
    \centering
	\includegraphics[width={500pt}, height=0.9\textheight]{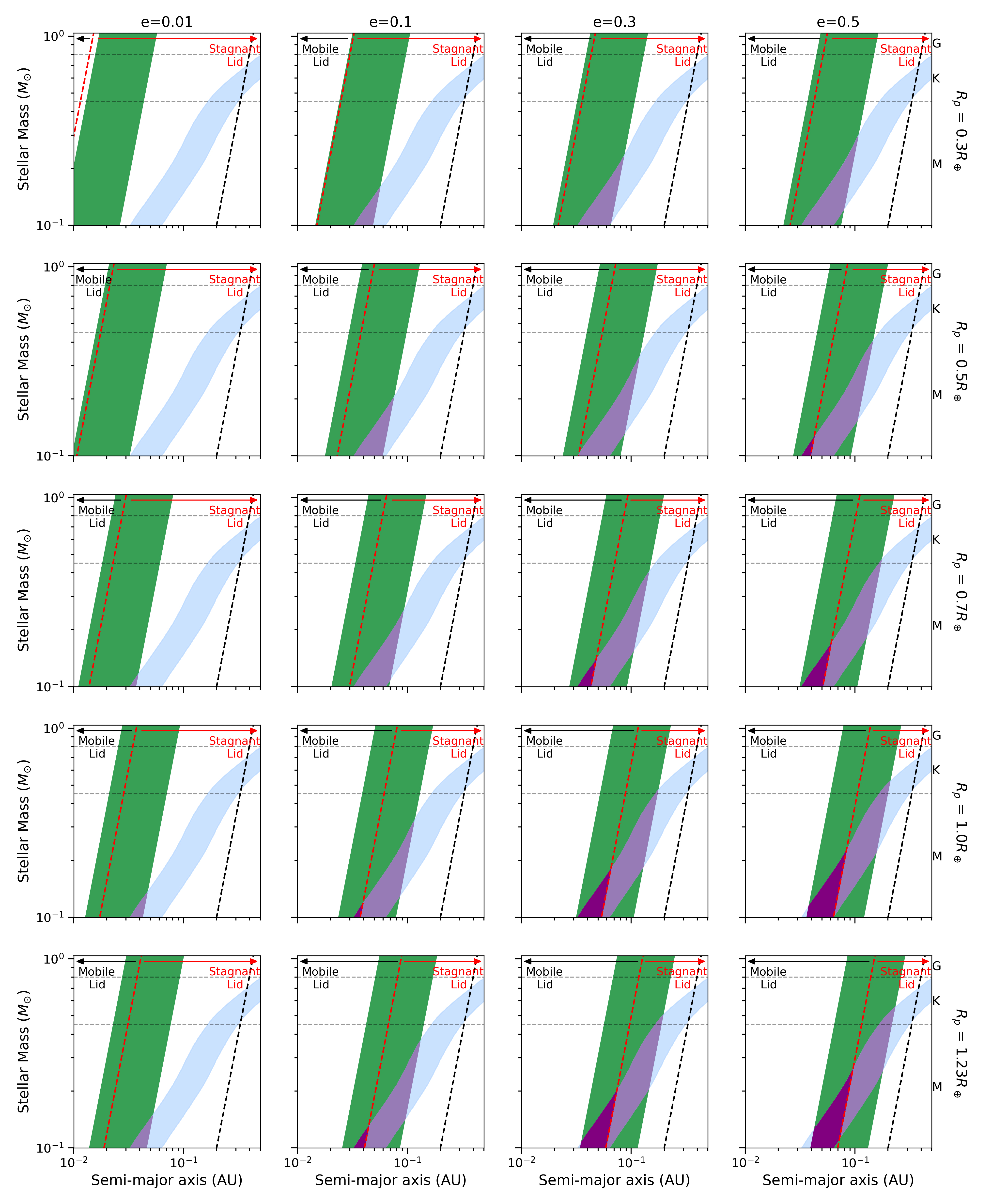}
    \caption{Modelled optimal tidal heating zones (green shaded regions) and tidal stress threshold (red dashed lines) against the CHZ (blue shaded regions), adjusting for parameters of planetary radius, eccentricity, semi-major axis, and stellar mass. Light purple regions represent the area of overlap between tidal heating zones and the CHZ. Dark purple shaded regions represent the area of overlap between tidal stress, tidal heating, and the CHZ. Black dashed lines represent the tidal locking radius from \citet{barnes2017tidal} Constant-Time-Lag model.} 
    \label{fig:figure4}
\end{figure*}

To expand the use of the information collated here, we simultaneously model the tidal stress and tidal locking thresholds as well as the optimal tidal heating zones in relation to the CHZ. In Figure~\ref{fig:figure4}, we display the effect that parameters of stellar mass, semi-major axis, eccentricity, and planetary radius have on tidal stress and tidal heating. We take the upper radius limit for rocky exoplanets as 1.23 R$_\oplus$ from \citet{chen2016probabilistic}, and a lower boundary of 0.3 R$_\oplus$, as the smallest exoplanet discovered around a main-sequence star – Kepler 37b – has a radius of $\sim$0.303 R$_\oplus$ \citep{haghighipour2015kepler}. For the eccentricity values, we model over a range from 0.01, which is a likely eccentricity estimate for tidally locked ultra-short period planets orbiting close to their host stars \citep{zanazzi2019ability}, to an eccentricity of 0.5, as the tidal stress and heating models change negligibly for \textit{e} > 0.5.

The first column in Figure~\ref{fig:figure4} presents the optimal tidal heating zones and tidal stress thresholds for an eccentricity \textit{e} = 0.01, spanning radii values from 0.3 R$_\oplus$ (first row) to 1.23 R$_\oplus$ (fifth row). In these graphs, we see that within the habitable zone there is no intersection between the tidal stress subduction threshold and the optimal tidal heating zone. Furthermore, only rocky exoplanets with radii between 0.7-1.23 R$_\oplus$ that reside within the inner edge of the habitable zone of low-mass M stars (M$_*$ < 0.15 M$_\odot$) have optimal tidal heating rates required for the maintenance of surface liquid water. With insufficient tidal stress and optimal tidal heating, this subset of rocky exoplanets, located within the light purple shaded regions in the first column of Figure~\ref{fig:figure4}, falls within the temperate stagnant lid tectonic regime, and requires vertical recycling, silicate weathering, or significant long-lived CO${}_{2}$ outgassing to support a stable climate suitable for the presence of surface liquid water \citep{tosi2017habitability,dorn2018outgassing,valencia2018habitability, foley2018carbon, foley2019habitability}. For R$_p$ = 0.3-0.5 R$_\oplus$ and \textit{e} = 0.01, there are no crossovers between tidal heating and the CHZ. Thus, all small exoplanets (R$_p$ < 0.5 R$_\oplus$) with an eccentricity of 0.01 residing within the CHZ would have terminal stagnant lid tectonics and would be unable to sustain surface liquid water unless supplemented with an additional source of heating such as radiogenic or primordial heating.

Increasing the eccentricity value to 0.1 in the second column of Figure~\ref{fig:figure4} corresponds to a general shift in the location of the optimal tidal heating regime so that tidally locked rocky exoplanets of all radii (R$_p$ = 0.3-1.23 R$_\oplus$) show a crossover between the CHZ and the optimal tidal heating zone enabling temperate stagnant lid tectonic activity, requiring additional mechanisms to support a stable climate and the presence of surface liquid water. Furthermore, for \textit{e} = 0.1, as denoted by the dark purple regions in column 2 of Figure~\ref{fig:figure4}, we see an overlap between tidal stress, tidal heating, and the CHZ for rocky planets with radii 1.0 R$_\oplus$ and 1.23 R$_\oplus$ orbiting in the inner CHZ (\textit{a} < 0.039AU and \textit{a} < 0.044AU) of low-mass M dwarfs with M$_*$ < 0.117 M$_\odot$ and M$_*$ < 0.13 M$_\odot$ respectively. Thus, it could be possible to have both mobile lid \textit{and} temperate tectonic activity conducive to the maintenance of surface liquid water on Earth-sized rocky exoplanets with \textit{e} = 0.1 orbiting the inner CHZ of low-mass M dwarfs.

In Figure~\ref{fig:figure4}, column 3, we can see that raising the eccentricity to 0.3 results in a shift of the optimal tidal heating regime into the CHZ of low-mass K stars. The \textit{e} = 0.3, R$_p$ = 1.23 R$_\oplus$ graph showcases the possibility of tidally driven temperate stagnant lid tectonic activity on rocky exoplanets located in the inner edge of the CHZ (\textit{a} < 0.193AU) of low-mass K stars (M$_*$ < 0.5 M$_\odot$). Additionally, when \textit{e} = 0.3, we see an overlap between tidal stress, tidal heating, and the CHZ for rocky exoplanets where R$_p$ = 0.7 R$_\oplus$, orbiting the CHZ at \textit{a} < 0.048AU of M stars with M$_*$ < 0.14 M$_\odot$. Therefore, increasing the eccentricity to 0.3 results in an increase in the range of radii where our tidal factors overlap with the CHZ, and temperate mobile tectonic activity conducive to the maintenance of surface liquid water could occur on rocky exoplanets with R$_p$ = 0.7-1.23 R$_\oplus$. 

Further increasing the eccentricity beyond 0.5 results in negligible changes, so the fourth column in Figure~\ref{fig:figure4} is approximately representative of exoplanets where \textit{e} $\ge$ 0.5. Exoplanets with the maximum eccentricity (\textit{e} = 0.5) and maximum radii (R$_p$ = 1.23 R$_\oplus$) could have temperate stagnant lid tectonic activity when orbiting the inner edge of the CHZ (\textit{a} < 0.229AU) of stars up to a maximum stellar mass of 0.55 M$_\odot$. Additionally, Earth-sized exoplanets (1.0 R$_\oplus$) with a high eccentricity value (\textit{e} = 0.5) fall within the temperate stagnant lid tectonic regime when orbiting the inner edge of the CHZ (\textit{a} < 0.206AU) of K stars (M$_*$ < 0.52 M$_\odot$). Therefore, the light purple shading in column 4 of Figure~\ref{fig:figure4} indicates an overlap between CHZ and the tidal heating regime, highlighting that it is possible for highly eccentric exoplanets orbiting K stars to have tidally driven temperate stagnant lid tectonic activity requiring additional mechanisms such as vertical recycling, silicate weathering, or significant long-lived CO${}_{2}$ outgassing to support a stable climate for the presence of surface liquid water  \citep{tosi2017habitability,dorn2018outgassing,valencia2018habitability, foley2018carbon, foley2019habitability}. When \textit{e} = 0.5, the overlap between the tidal stress threshold, optimal tidal heating regime, and the CHZ now includes exoplanets of radii 0.5 R$_\oplus$ orbiting the inner edge of the CHZ (\textit{a} < 0.042AU) of low-mass M dwarfs (M$_*$ < 0.125 M$_\odot$). Thus, highly eccentric tidally locked rocky exoplanets where R$_p$ = 0.5-1.23 R$_\oplus$ could be able to sustain tidally induced temperate mobile lid tectonic activity, that is potentially comparable to plate tectonics on Earth, which would help in maintaining the presence of surface liquid water. However, as seen in the first row of Figure~\ref{fig:figure4}, regardless of eccentricity, small rocky exoplanets of radii 0.3 R$_\oplus$ would be unable to experience significant stresses to aid or initiate temperate mobile lid tectonics in the CHZ. Finally, the subset of highly eccentric (\textit{e} = 0.5) rocky exoplanets with R$_p$ = 1.0-1.23 R$_\oplus$ orbiting the CHZ at \textit{a} < 0.038 and \textit{a} < 0.043AU respectively would have tidal heating rates above 300 $Wm^{-2}$, indicating a high probability of greenhouse runaway effect desiccating the exoplanet. For these hot mobile lid exoplanets, the addition of radiogenic or primordial heating would only increase the heating rate to even more inhospitable levels, so these exoplanets are unlikely to be able to retain surface liquid water.

\section{Discussion} \label{sec:disc}

Here, we are exclusively estimating tidal heating, as this is predicted to be the primary source of internal heating for tidally locked rocky exoplanets. Other sources such as primordial or radiogenic heating could also contribute to the exoplanet’s heating rate. However, estimating an exoplanet’s primordial and radiogenic heating rates depends on numerous factors for which we do not currently have sufficient information, such as the internal composition and structure of the planet, and is beyond the scope of this paper \citep{dorn2018assessing}. Nevertheless, radiogenic and primordial heating would only add to the internal heating budget. Therefore, as we can see from Figure~\ref{fig:figure2}, primordial and radiogenic heating could only make a positive contribution to those terminal stagnant lid exoplanets that have low tidal heating rates, as with the addition of extra internal heating sources these exoplanets could shift into the temperate stagnant lid category.

The modelling employed in this paper provides a general approximation of tidal stress, tidal heating rates and tectonic activity for known exoplanets; however, it is important to acknowledge its simplicity. Here, we utilise an idealised model of tidal stress, which assumes a constant density incompressible exoplanet, with shear modulus and viscosity values based on the pressure and temperature of rocky planets found in our solar system. Due to the limitations of knowledge regarding interior structures and compositions of exoplanets, \citet{zanazzi2019ability} estimate that these assumptions make the tidal stress calculations uncertain beyond an order of magnitude estimate. Additionally, we follow the assumption from \citet{jackson2008tidal} that terms of higher order than ${e}^2$ are negligible. While tidal heating models where these higher-order corrections are incorporated do exist and could significantly affect heating rates, they also involve numerous assumptions regarding planetary dissipation processes and require parameters that we currently have no measured values for \citep{mardling2002calculating,efroimsky2012tidal,matsuyama2017tidal}.

Through these simplified models, we are attempting to make use of the limited observational data on the characteristics of rocky exoplanets we currently have available. The tidal heating and tidal stress models utilised here are exclusively relevant for tidally locked exoplanets, which comprise $\sim$99\% of currently observed rocky exoplanets \citep{mcintyre2019planetary}; thus, the models are not applicable for the remaining 1\% of exoplanets with non-synchronous rotation. Furthermore, there will be a time pre-dating tidal locking where the non-synchronous rotation could have induced stronger stresses and potentially initiated mobile lid tectonic activity.

While tidal stress and tidal heating rates are predicted to have a significant impact on the initiation and maintenance of tectonic activity on tidally locked rocky exoplanets, other factors such as climate, gravity, and the presence of an initial water budget are additional mechanisms that could weaken the lithosphere and reduce the yield strength of a plate, leading to an increased possibility of subduction and mobile lid tectonic activity \citep{valencia2007inevitability,o2007conditions,korenaga2010likelihood,tikoo2017fate,stern2018stagnant}. As more information on the climate and existence of liquid water on the surface of rocky exoplanets becomes available, future studies will be able to determine whether the presence of gravity and surface liquid water are the main drivers in weakening an exoplanet’s lithosphere and causing mobile lid tectonic activity, or whether tidal stress and tidal heating alone could be sufficient to drive mantle convection. 

In this paper, we assume that due to close proximity to their host star, tides work to drive the planet's rotational period toward synchronization with their orbital period, and thus, when the exoplanets are tidally locked, they are also rotating synchronously. However, exoplanets within our sample could escape synchronous rotation by being captured in spin-orbit resonances \citep{goldreich1966spin, makarov2012dynamical, rodriguez2012spin}, and resonant planet-planet interactions with their exterior planetary companions \citep{delisle2017spin, vinson2017spin, zanazzi2019ability}. According to \citet{henning2020tidal}, the introduction of non-synchronous rotations could significantly reduce tidal heating habitability widths. Additionally, these tidal models do not account for obliquity and, as \citet{wang2016effects} results suggest, with higher obliquities, the habitability of M dwarfs narrows. Finally, our understanding of tectonic activity on exoplanets is limited and requires further exploration into the differentiation of the boundary between mobile and stagnant lid regimes \citep{bercovici20157}. 

In the future, with more spectroscopic information from upcoming space-based missions, providing additional atmospheric information on planets, we will be able to make stronger conclusions regarding lithospheric stresses, internal heating rates, subsequent tectonic activity, and the maintenance of long-term surface liquid water.

\section{Conclusion} \label{sec:concl}

This paper provides a starting point for the application of the limited amount of observational data to the postulate that tectonic activity driven by tidal stress and heating could play an important role in establishing suitable conditions for the maintenance of surface liquid water. An analysis of tidal stress needed to aid mantle convection and initiate subduction reveals that the majority of exoplanets in our sample ($\sim$70\%) have a stagnant lid tectonic regime. Subsequent examination of tidal heating rates suggests that 63\% $\pm$ 2\% of the 767 exoplanets from our sample would currently fall within the internal heating requirements to support a stable climate. Finally, 10\% $\pm$ 1\% of exoplanets in our sample both reside within the optimal tidal heating regime \textit{and} pass the subduction threshold. This subset of exoplanets could be able to sustain tidally induced temperate mobile lid tectonic activity comparable to plate tectonics on Earth, that would help in maintaining the presence of surface liquid water. Furthermore, when examining the CHZ subset of our sample, 57\% $\pm$ 5\% of tidally locked rocky CHZ exoplanets reside within the optimal tidal heating range between 0.04 and 300 $Wm^{-2}$. Of these temperate exoplanets, Proxima Cen b and GJ 1061 d have subduction values indicative of mobile lid tectonic activity that could help in maintaining the presence of surface liquid water. Additionally, 43\% $\pm$ 5\% of CHZ exoplanets reside within the optimal tidal heating regime; however, their  subduction values are indicative of stagnant lid tectonic activity, requiring additional mechanisms such as vertical recycling, silicate weathering, or significant long-lived CO$_2$ outgassing to sustain stable climates suitable for the presence of surface liquid water \citep{tosi2017habitability,dorn2018outgassing,valencia2018habitability, foley2018carbon, foley2019habitability}.

Finally, when broadening our modelling to determine the intersection between tidal stress, tidal heating, and the CHZ to discover optimal regions to target for future observations, we conclude that tidally driven temperate tectonic activity conducive to the maintenance of surface liquid water occurs predominantly around M dwarfs, and with these stars accounting for $\sim$70\% of the local stellar population future observations of exoplanets orbiting these stars will be important for further analysis of tidally driven tectonic activity \citep{kochukhov2021magnetic,winters2019solar}. We also find that, independent of eccentricity and radius, tidally locked rocky planets orbiting in the CHZ at \textit{a} > 0.229AU around stars larger than 0.55 M$_\odot$ are terminal stagnant lid exoplanets and would be unable to sustain temperate tectonic activity via tidal factors alone, requiring an additional form of internal heating such as primordial or radiogenic to initiate and maintain tectonic activity suitable for the presence of surface liquid water. Furthermore, modelling from Figure~\ref{fig:figure4} shows that there exist intersections between the tidal heating zone and the tidal stress threshold on Earth-sized (R$_p$ = 1.0-1.23 R$_\oplus$), eccentric (\textit{e} > 0.1) exoplanets orbiting in the CHZ of low-mass M dwarfs, where both mobile lid \textit{and} optimal tidal heating could be sustained. This subset of exoplanets could display tidally driven temperate mobile lid tectonic activity comparable to plate tectonics on Earth.

The results presented here highlight the fact that despite satisfying the classical definition of habitability, $\sim$40\% of the tidally locked rocky exoplanets from our sample located within the CHZ would be unable to support the temperate tectonic activity needed for a stable climate and thus are unlikely to retain surface liquid water. This further emphasises the need for the habitability of each exoplanet to be assessed individually, while continuing to expand the definition of what makes a planet habitable. Tidal stress and tidal heating are amongst many astronomical and geophysical factors that could affect an exoplanet's habitability and its ability to sustain surface liquid water. Increasing our knowledge and information on planets, stars, and their interactions over time will help in analysing and evaluating additional habitability parameters that we can use to optimise target selection for future observations further characterising exoplanets and their atmospheres. 

\begin{acknowledgements}
I thank Rory Barnes for refereeing this paper and providing informative comments that have improved the quality and clarity of this work. I also thank Brian Jackson for the helpful discussion and advice. This research was supported by an Australian Government Research Training Program (RTP) Scholarship.

\end{acknowledgements}

\bibliographystyle{aa}
\bibliography{srnmc}
\newpage
\onecolumn

\begin{landscape}
\begin{appendix}

\section{Data Tables} 
\tiny

\begin{TableNotes}
\footnotesize
\item \text{*} Stars with no age estimates for which we assume a lower age limit of 1 Gyr 
\item $^\wedge$ Planets located in the optimistic CHZ as defined by \citep{kopparapu2013habitable,kopparapu2014habitable}
\end{TableNotes}

\begin{longtable}{lcccccccc}
\caption{Rocky Exoplanets Sample (in order of increasing tidal heating rate)}\\
\label{tab:exostres}\\
\hline\hline
Planet Name & Planet Mass (M$_\oplus$) & Planet Radius (R$_\oplus$) & Stellar Mass (M$_\odot$) & Eccentricity & Semi-major Axis (AU) & Stellar Age (Gyr) & ${h}$ (10${}^{5}$) & ${H}$ (10${}^{5}$)\\
\hline
\endfirsthead
\caption{Continued.} \\
\hline
Planet Name & Planet Mass (M$_\oplus$) & Planet Radius (R$_\oplus$) & Stellar Mass (M$_\odot$) & Eccentricity & Semi-major Axis (AU) &  Stellar Age (Gyr) & ${h}$ (10${}^{5}$) & ${H}$ (10${}^{5}$)\\
\hline
\endhead 
\hline
\endfoot
\hline
\insertTableNotes
\endlastfoot
K6889.01 & 0.12  $\pm$ 0.04 & 0.55 $\pm$ 0.06 & 1.02 $\pm$ 0.07 & 0.01 $\pm$ 0.01 & 0.354 $\pm$ 0.008 & 3.7 & 0.001 $\pm$ 0.001 & 0.0000003 $\pm$ 0.0000001 \\
K7673.01 & 0.43  $\pm$ 0.23 & 0.79 $\pm$ 0.11 & 0.74 $\pm$ 0.03 & 0.01 $\pm$ 0.01 & 0.331 $\pm$ 0.005 & 7.9 & 0.001 $\pm$ 0.001 & 0.0000006 $\pm$ 0.0000003 \\
K5704.01 & 1.98  $\pm$ 0.80 & 1.21 $\pm$ 0.13 & 0.72 $\pm$ 0.03 & 0.01 $\pm$ 0.01 & 0.368 $\pm$ 0.005 & 14.9 & 0.001 $\pm$ 0.001 & 0.0000009 $\pm$ 0.0000003 \\
K3061.02 & 1.04  $\pm$ 0.32 & 1.01 $\pm$ 0.09 & 0.78 $\pm$ 0.04 & 0.01 $\pm$ 0.01 & 0.328 $\pm$ 0.006 & 6.7 & 0.001 $\pm$ 0.001 & 0.0000015 $\pm$ 0.0000005 \\
K5954.01 & 1.00  $\pm$ 0.45 & 1.00 $\pm$ 0.12 & 0.98 $\pm$ 0.06 & 0.01 $\pm$ 0.01 & 0.343 $\pm$ 0.007 & 4.6 & 0.001 $\pm$ 0.001 & 0.0000018 $\pm$ 0.0000008 \\
K8076.01 & 1.60  $\pm$ 0.41 & 1.14 $\pm$ 0.08 & 0.92 $\pm$ 0.06 & 0.01 $\pm$ 0.01 & 0.327 $\pm$ 0.007 & 4.8 & 0.002 $\pm$ 0.001 & 0.000003 $\pm$ 0.000001 \\
Kepler-409 b & 1.55  $\pm$ 0.84 & 1.13 $\pm$ 0.16 & 0.91 $\pm$ 0.05 & 0.01 $\pm$ 0.01 & 0.319 $\pm$ 0.006 & 5.8 & 0.002 $\pm$ 0.001 & 0.000004 $\pm$ 0.000002 \\
K5213.01 & 0.25  $\pm$ 0.23 & 0.68 $\pm$ 0.15 & 0.90 $\pm$ 0.05 & 0.01 $\pm$ 0.01 & 0.251 $\pm$ 0.004 & 4.4 & 0.001 $\pm$ 0.001 & 0.000005 $\pm$ 0.000004 \\
\hline
\end{longtable}

\end{appendix}
\end{landscape}
\end{document}